\documentclass{jetpl}
\twocolumn
\usepackage{epsfig}
\lat

\title{Indication of  asymptotic  scaling
in the  reactions  $dd\to p^3$H, $dd\to n^3$He  
  and $pd\to pd$  }

\rtitle{Indication of  asymptotic  scaling    }

\sodtitle{ Indication of  asymptotic  scaling  
 in the  reactions  $dd\to p^3$H 
  and $dp\to dp$  }

\author{Yu.\,N.\,Uzikov$^{+}$ \/\thanks{e-mail: uzikov@nusun.jinr.dubna.su }}

\address{$^+$Joint Institute for Nuclear Research, LNP, Dubna, Moscow
 region 141980, Russia\\~\\
}

\dates{11 January 2005}{28 February 2005}
    
 \abstract{
 It is shown that the differential cross sections of the reactions
 $dd\to~^3He\,n$ and $dd\to~^3H\,p$ measured at c.m.s. scattering
 angle  $\theta_{cm}=60^\circ$ in the interval of the deuteron beam energy
 0.5 - 1.2 GeV demonstrate the scaling behaviour,
 $d\sigma/d t\sim s^{-22}$, which follows from  constituent quark counting
 rules. It is found  also that  the differential cross section
 of the elastic $dp\to dp$ scattering at $\theta_{cm}=125^\circ-135^\circ$
  follows  the scaling regime $\sim s^{-16}$ at  beam energies
 0.5 - 5\, GeV. These data are parameterized  here using the
 Reggeon exchange. 
 }

\begin{document}
\maketitle



 Nuclei and nuclear reactions at low and intermediate energies
 (or at long and medium distances between nucleons $r_{NN}>0.5 fm$)
 traditionally are
 described in terms of effective nucleon-nucleon interactions which are
 mediated by the exchange of mesons. 
 In the limit of
 very high energies ($s\to \infty$) and
 transferred four-momenta ($t\to \infty$)  the  perturbative quantum
 chromodynamics (pQCD) is expected to be applied for  explanation of
 nuclear reactions in terms of  quarks and gluons.    
 At present, one of the most interesting problems 
 in  nuclear physics
 is an interplay between the meson-baryon and quark-gluon pictures of
 the strong  interaction.
 The main  question  is the following:
 at which  $s$ and $t$ values
 (or, more precisely, relative  momenta $q$ of nucleons in
  nuclei) does   the transition region
 from the meson-baryon to the quark-gluon picture of nuclei set in?

A possible signature for this transition is given 
by the  constituent
 counting rules (CCR) \cite{matveevmt,brodsky73}. According to
 dimensional  scaling \cite{matveevmt,brodsky73} and
  pQCD \cite{brodsky75},  the differential cross
 section  of a binary reaction $AB\to CD$ at  high enough
 incident energy  can be parameterized for a given c.m.s. scattering angle
 $\theta_{cm}$  as
\begin{equation}
\label{general}
\frac{d\sigma}{d\,t}(AB\to CD)= \frac{f(t/s)}{s^{n-2}},
\end{equation}
 where 
 $n=N_A+N_B+N_C+N_D$ and $N_i$ is the minimum number of 
 point-like constituents in the {\it i}th hadron (for a lepton one has
 $N_l=1$),
 $f(s/t)$ is a function of  $\theta_{cm}$.
 Existing data  for 
 many  measured  hard scattering processes
 with free hadrons appear to be consistent with Eq. (\ref{general})
 \cite{andersen76}.  
   At present, in a nuclei sector  only electromagnetic processes on
 the deuteron were found to be compatible  with the CCR.
 So, the deuteron electromagnetic 
 form factor measured  at  SLAC \cite{arnold} and JLab 
  \cite{alexa} at high momentum transfer  $Q^2>4$ GeV$^2$ approaches
 to  the scaling as $\sqrt{A(Q^2)}\to Q^{-10}$
 what is in agreement with the CCR.
 The deuteron two-body photodisintegration cross section $\gamma d\to pn$
 demonstrates   the $s^{-11}$ scaling behaviour in the data  obtained in SLAC  
\cite{napolitano,freedman,beltz}
 at $E_\gamma> 1$ GeV, $\theta_{cm}\approx  90^\circ$ and Jlab \cite{bochna} 
at $E_\gamma=1-2$ GeV for $\theta_{cm} =89^\circ, 69^\circ$ \cite{schulte}.
 According to the data
 \cite{bochna}, at photon energy 3.1 GeV  and scattering angle
 $\theta_{cm}=36^\circ$ there is no evidence  for the $s^{-11}$ scaling.  
 A nearly complete angular distribution of the cross section
 of this reaction recently measured at energies $0.5 - 3.0$ GeV
 \cite{mizarata} demonstrates
 the $s^{-11}$ behaviour at proton transverse momentum $p_T>1.1$ GeV/c
 \cite{rossi}. 
 Meson-exchange models fail 
 to explain   the 
 $\gamma d\to pn$  data at $E_\gamma >1$ GeV (see, for example,
\cite{bochna} and references [3],[4],[9], [10] therein).
 Recent models based
 on quark degrees of freedom  have  become quite successful
 in describing this data. 
 Thus, the observed in Ref. \cite{freedman,schulte}
 forward-backward asymmetry
 was described within the Quark-Gluon String 
(QGS) model \cite{grishina} using  a non-linear
 Regge  trajectory of the nucleon. Other quark models 
 applied to this reaction 
 are reviewed in Ref. \cite{gillmangross}.

 The dimensional scaling was derived before 
 the QCD was discovered. The main  assumption was an automodellism
 hypothesis  for the amplitude of the binary  reaction
  with point-like constituents in colliding (and outgoing)
 particles and high  enough $s$ and $t$
\cite{matveevmt}. The pQCD (and, consequently,
 the scaling behaviour within the pQCD)
 is expected to be valid  at very high
 transferred momenta  which are not yet reached in 
 existing data for nucleon and deuteron form factors \cite{isgur,farrar}.
 From this point of view the origin of the scaling behaviour observed
  in the reactions
 with the deuteron at moderate transferred momenta
 \cite{arnold,alexa,napolitano,freedman,beltz,bochna,schulte,mizarata}
  is unclear and  considered in some papers as a potentially
  misleading indicator of the success
  of pQCD \cite{gillmangross}. Moreover, the hadron helicity conservation
  predicted  by the pQCD was not confirmed experimentally  in the scaling
  region (see Ref. \cite{wijesoor} and references therein).  
 On the other side, in these reactions the
 3-momentum transfer $Q=1-5$ GeV/c  is  large enough  to probe 
 very  short  distances between nucleons   in nuclei,
 $r_{NN}\sim 1/Q<0.3 \,fm$,
 where $0.3\, fm$ is a size of a constituent quark
\cite{diakonov}.
  One may expect that nucleons lose their separate identity
in this overlapping region and, therefore, six-quark
(or, in general case, multi-quark) components of a nucleus can
 be probed in these reactions.
 In order to get more insight into  the underlying dynamics of the
 scaling behaviour new data are necessary, in particular, for
 hadron-nuclei interactions. 
 
\begin{figure}[hbt]
\mbox{\epsfig{figure=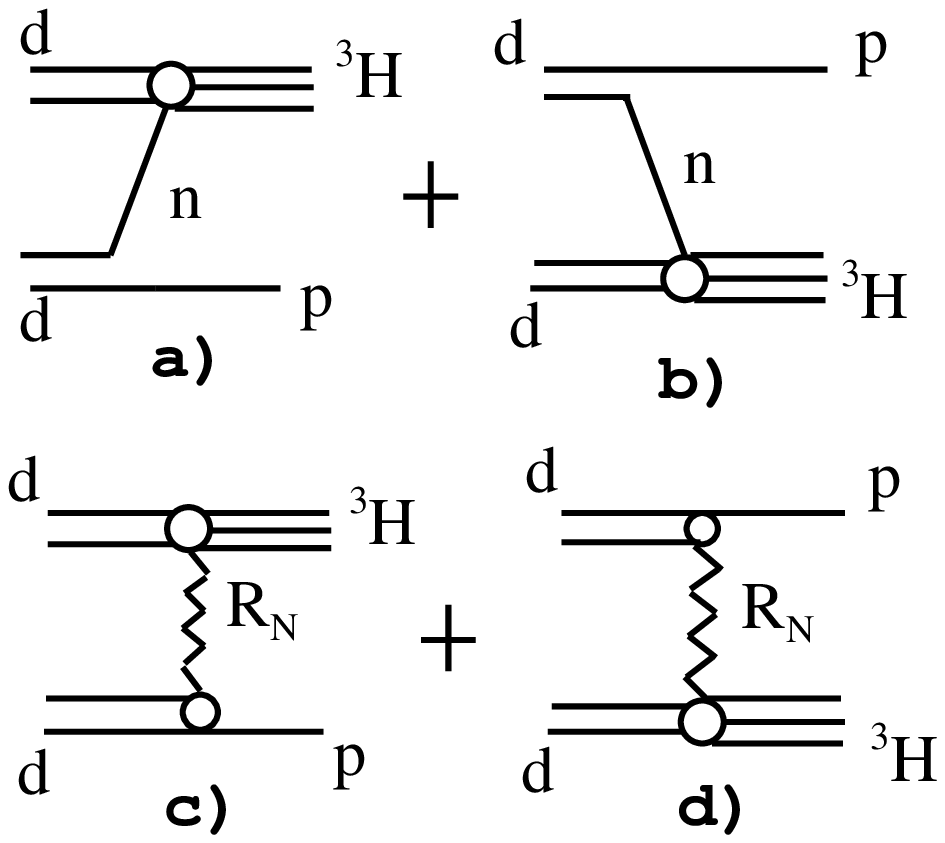,height=0.5\textheight, clip=}}
\caption{ The mechanisms of the reaction $dd\to ~^3H\,p$:
  one nucleon exchange (a-b),  Reggeon exchange (c-d). 
 }
\label{one}
\end{figure}

\begin{figure}[hbt]
\mbox{\epsfig{figure=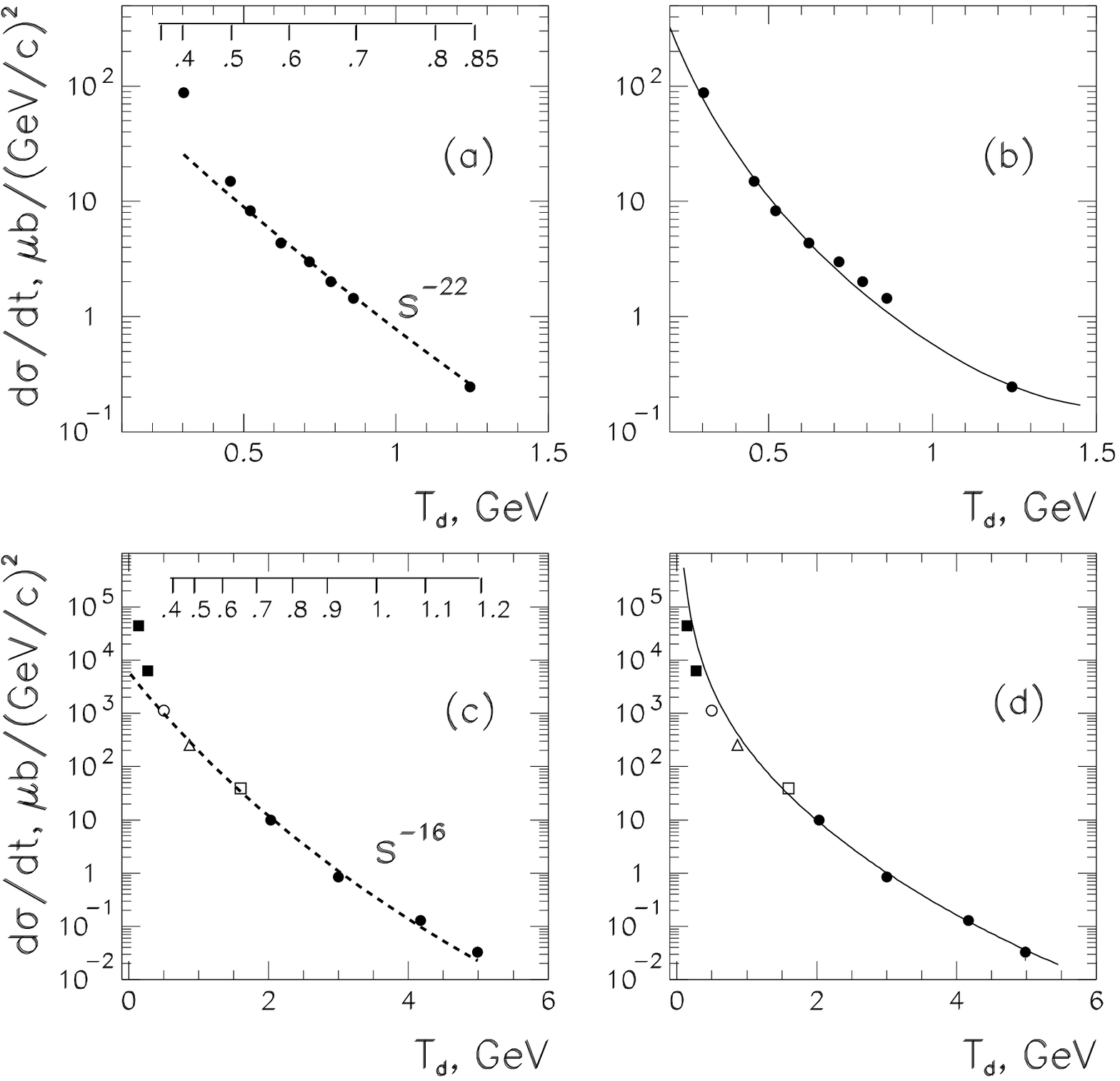,height=0.7\textheight, clip=}}
\caption
{The differential 
 cross section of the $dd \to~^3He\,n$
 and $dd \to~^3H\,p$  reactions at $\theta_{c.m.}=60^o$  (a,\, b)
 and $dp \to dp$  at $\theta_{c.m.}=127^o $ (c,\, d)
 versus the deuteron beam kinetic energy.
 Experimental data in  (a,\, b)  are
 taken from \protect\cite{bizard}.
 In  (c,\, d), the  experimental data (black squares),($\circ$), ($\triangle$),
(open square) and
($\bullet $) are taken from \protect\cite{sekiguchi},
\protect\cite{hatanaka}, \protect\cite{booth}, \protect\cite{winkelman}
 and  \protect\cite{dubal}, respectively. 
  The dashed curves give the $s^{-22}$ (a) and $s^{-16}$ (c) behaviour.
  The full curves show the result 
  of calculations using Regge formalism 
  given by Eqs. (\ref{sumab}),
 (\ref{tamplituda}), (\ref{vichet})  with the
  following parameters:
  (b) -- $C_1= 1.9\, GeV^2,\, R_1^2=0.2\, GeV^{-2}, C_2= 3.5,\, \,
  R_2^2=-0.1\, GeV^{-2}$; 
   (d) -- $C_1 =7.2\, GeV^2,\, R_1^2=0.5\, GeV^{-2}$, 
$ C_2=1.8,$\, 
$ R_2^2=-0.1\, GeV^{-2}$.
  The upper scales in (a) and (c)  show the relative momentum $q_{pn}$ 
(GeV/c) in the deuteron for the ONE mechanism.
}
\label{fig2}
\end{figure}

  In the present paper we show   that in hadron
 interactions with participation of the lightest
 nuclei $^2H$, $^3H$ and  $^3He$ 
 the scaling behaviour given by
 Eq. (\ref{general}) is also occurs, specifically, at 
 beam energies around 1 GeV  if the scattering angle   is  large enough.
  In order to estimate at which internal momenta $q_{pn}$
 between nucleons  in the deuteron one should expect the scaling onset, we
 consider here the reaction $\gamma d\to pn$ assuming that the  one nucleon 
 exchange (ONE) mechanism dominates. Under this assumption the cross section
 is proportional to the squared wave function of the deuteron in momentum space
 $d\sigma\sim |\psi_d(q_{pn})|^2$. Using relativistic kinematics we  obtain
 that $q_{pn}$ is
 larger than 1\,GeV/c
 at the photon energy $E_\gamma>1$  GeV and 
$\theta_{cm}=90^\circ$.  Furthermore, assuming  for
 the reaction $dd\to~^3Hp$ (or
 $dd\to~^3Hen$) that the ONE mechanism dominates (Fig.\ref{one},a-b), 
 we  get  $d\sigma\sim |\psi_d(q_{pn})|^2\times
|\Psi_h(q_{Nd})|^2$, where $\Psi_h(q_{Nd})$ is the overlap between the
$^3H(~^3He)$ and deuteron wave functions and $q_{Nd}$ is the $N-d$ relative
 momentum in the $^3H(~^3He)$.  On this basis we obtain, 
 for example, at $T_d= 0.8$ GeV and
 $\theta_{cm}=90^\circ$  the relative momenta 
 $q_{pn}=0.8$ GeV/c and $q_{Nd}=1.0$ GeV/c. This values are close 
 to those we have found  for
  the $\gamma d\to pn$ reaction in the scaling region.
 Therefore
 one may expect that the scaling behaviour  in the $dd\to ~^3He\,n$ reaction 
 occurs in the  GeV region for large scattering angles,
 $\theta_{cm}\sim 90^\circ$.
 In Fig.\ref{fig2}a,b we show the experimental data from Ref.\cite{bizard}
 obtained at SATURNE at beam 
 energies 0.3 - 1.25 GeV for the maximum measured
 scattering  angle $\theta_{cm}=60^\circ$. Shown on the upper scale is 
 the minimum relative momentum  in the deuteron for the ONE diagram.  
 One can see that at beam energies $0.5-1.25$ GeV the data
  perfectly follow  the $s^{-22}$ dependence. (In this reaction
 $n=6+6+9+3=24$).
 In Fig. \ref{fig2}a the dashed curve
 represents the $s^{-22}$ dependence with arbitrary normalization fitted
 on the data  with $\chi_{n.d.f.}^2= 1.18$. 
 For the ONE  diagram  in Fig. \ref{one} b, which dominates at
 $\theta_{cm}=60^\circ$,
 this region corresponds to the internal momenta
 $q_{pn}=$0.55\,GeV/c - 0.85\,GeV/c  in the deuteron
  and $q_{Nd}=$  0.60 GeV/c - 0.94 \,GeV/c
 in the $^3He$ ($^3H$)  nuclei. Therefore, within
 this model the probed  NN-distances in the deuteron are less  than 
 $r_{NN}< 1/0.55$ GeV/c= 0.35\, fm. This regime, in principle, corresponds
 to  formation of a six quark configuration in the deuteron. 
  At $\theta_{cm}=90^\circ$
 the diagrams in Fig.\ref{one}a and b  are equivalent and correspond
 to higher  momenta $q_{pn}=$ 0.7 - 1.1\,GeV/c and $q_{Nd}=$ 0.80 -1.22
 \,GeV/c for the same  beam energies  0.5 - 1.25 GeV.
 Therefore,
 continuation of measurements up to $\theta_{cm}=90^\circ$ is very desirable
 to confirm the observed $s^{-22}$ behaviour.
 One can see in linear scale that the cross section $s^{-22} d\sigma/dt$,
 as a  function of $T_d$,  demonstrates some oscillations 
  which are similar to those observed in pp-scattering at
  $\theta_{cm}=90^\circ$ \cite{pire}.
  However, the number of available  experimenatal points is too small in
  the  scaling region  (5 or 6) to make a definite
  conclusion.

  The $dp\to dp$ data  obtained
 in different experiments \cite{sekiguchi,hatanaka,booth,winkelman,dubal}
 at the c.m.s. scattering angle $\theta_{cm}=127^\circ$ are shown 
 in Fig. \ref{fig2}c,d 
 versus the deuteron beam energy $T_d$. This scattering angle corresponds to
 a region of the minimum in the angular dependence of the differential
 cross section $dp\to dp$, where
 the contribution of the three-body forces (and non-nucleon degrees of freedom
 in the deuteron) is expected to be best pronounced
\cite{witala98,uzikov}.  
 One can see that at low energies ($<0.25$ GeV)
 the cross section falls very fast with increasing $T_d$, but
 the slope of the energy dependence is sharply changed  at about 0.5 GeV. 
 Above this energy the cross section is appeared to follow 
 the  $s^{-16}$ scaling behaviour. (In the $dp\to dp$ one has
 $n=3+6+3+6=18$). We can show that a similar behaviour is observed at
 $\theta_{cm}=135^\circ$.
 However, the parameter
 $\chi_{n.d.f.}^2$ is rather high for the $dp\to dp$ data,
 $\chi_{n.d.f.}^2 = 4.3$. The high 
 $\chi_{n.d.f.}^2$ value  can be, probably, addressed to
 uncertainties in systematic 
 errors which are different in various experiments  \cite{dubal}.
 Therefore,  new, more detailed data, are requested preferably from one 
  experiment  covering the whole interval of 
  energies $T_d=1-5$ GeV.
 We notice that the discrepancy observed  in \cite{witala2004}
 between the results of the Faddeev calculations and the measured 
 unpolarized cross section of the $pd\to pd$
 at $T_p=0.25$ GeV (corresponding to $T_d=0.5$ GeV in the $dp\to dp$),
 is, presumably,  caused by the deuteron  six-quark
 component which is  not taken into account in \cite{witala2004} but,
 as seen  from Fig. \ref{fig2}c,  starts playing in the $pd\to pd$
 at this kinematics.   
  
  Due to very high internal momenta in the $d\leftrightarrows
 pn$ and $Nd\leftrightarrows ~^3H(^3He)$
 vertices, $q \sim 1$\, GeV/c, calculation with the $^3H(~^3He)$ and
 deuteron wave  functions 
 obtained from the Schr\"odinger equation
 with conventional NN-potentials are likely unrealistic.
 Since in the reactions $\gamma d\to pn$,\, $dd\to~^3H\,p$
 (or $dd\to~^3He\,n$) and $dp\to dp $ (in  the backward hemisphere)
 an important contribution comes from the baryon exchange mechanism,
 for numerical estimations we apply here  the Reggeon exchange
 formalism  developed earlier for the $pp\to d\pi^+$
 reaction at $-t<1.6$ (GeV/c)$^2$ \cite{kaidalov} and the  $\gamma d\to pn$
 at $E_\gamma >1 $ GeV \cite{grishina}.
 In this way one may estimate 
 to what extent the observed scaling
 behaviour in the $dd\to~^3Hen$   ($dd\to~^3Hp$) and $dp\to dp$ reactions
 is connected to that in the $\gamma d\to pn$.
 The amplitude of the reaction $dd\to~^3H\,p$
can be written as
\begin{equation}
\label{sumab}
T=T(s,t)+T(s,u),
\end{equation}
 where the first (second)  term corresponds to the diagram in Fig.
 \ref{one}~{a} ({b}) and
 the sign plus is chosen due to the Bose-statistics for the deuterons.
The amplitude $T(s,t)$ is written in the Regge form:
\begin{equation}
\label{tamplituda}
T(s,t)= F(t)\left (\frac{s}{s_0} \right )^{\alpha_N(t)}
\exp{\left [ -\frac{i \pi}{2}\left (\alpha_N(t)-\frac{1}{2}\right ) \right ]}.
\end{equation}

 We  use here
 the effective  Regge trajectory for the nucleon from \cite{kaidalov}:
 $\alpha_N(t)=\alpha_N(0)+
\alpha_N^\prime\,t + \alpha_N^{\prime \prime}/2 \, t^2$
 with the parameters $\alpha_N(0)=-0.5$, $\alpha_N^\prime=0.9$\, GeV$^{-2}$
 and $\alpha_N^{\prime \prime}=0.4$\, GeV$^{-4}$, so 
$\alpha_N(m_N^2)=\frac{1}{2}$, where $m_N$ is the nucleon mass. 
 The function  $F(t)$ is parameterized as \cite{kaidalov}
\begin{equation}
\label{vichet}
F(t)=\frac{C_1\exp{(R_1^2\,t)}}{m_N^2-t}+C_2\exp{(R^2_2t)},
\end{equation} 
where  the first term explicitely takes into account the nucleon pole
in the $t$ channel. According to \cite{kaidalov}, the second term 
at $R^2\approx 0$  is important
 at $|t|>1$ GeV$^2$, that indicates to a presence of  structureless
 configurations  in the deuteron ($^3He$, $^3H$) wave functions
 at short distances.   The results of calculation  are shown
 in Fig. \ref{fig2}~{b} 
and parameters $C$ and $R^2$ are given in the caption. 
 One can see a fairly good agreement with the data. 
 For the reactions $dd\to~^3H\,p$ and $dd\to ~^3He\,n$ 
 the parameter $R_1^2$ is lower  in comparison
 with that used in  \cite{kaidalov}
($R_1^2=3$\, GeV$^2$)
 to fit  the $pp\to d\pi^+$ data.
 Such a diminishing
  $R_1^2$ is likely  connected to  a much   more intensive
 high momentum nucleon component of the $^3He(~^3H)$ wave function 
 as compared  with the deuteron \cite{pandaripande}.
  The increasing ratio $C_1/C_2$
  could mean that multiquark configurations in the
 $^3He(~^3H)$  become more  important at given $t$ as compared with
  the deuteron. We also performed this analysis for  
 the  $dp\to dp$ reaction and obtained  a good agreement with the data
 under minor modification of the parameter $R_1^2$ and $C_1/C_2$
 (see Fig. \ref{fig2}{ d}).

  In conclusion, 
 the CCR scaling behaviour is observed
 in  cross sections of  hadron-nucleus reactions  with the deuteron and
 $^3He\, (~^3H$) nuclei. This behaviour sets in at energies around
 1 GeV and large scattering angles, where  high momentum components
 of nuclear wave functions are required in  the Schr\"odinger formalism. 
 To confirm this observation,   
 more detailed data are  necessary for these and other exclusive reactions
 in the $pd$, $dd$, $p^3He$ collisions, probably,  including meson production.

  I am thankful to V.I.~ Komarov, A.V.~ Kulikov, V.V.~Kurbatov, B.~Pire  and
 H.~Seyfarth  for stimulating discussions.



 \eject 
\onecolumn
\eject 
\end{document}